\documentclass{article}
\usepackage{ijcai24}
\usepackage{times}
\usepackage{soul}
\usepackage{url}
\usepackage[hidelinks]{hyperref}
\usepackage[utf8]{inputenc}
\usepackage[small]{caption}
\usepackage{graphicx}
\usepackage{amsmath}
\usepackage{amsthm}
\usepackage{booktabs}
\usepackage{algorithm}
\usepackage{algorithmic}
\usepackage[switch]{lineno}
\usepackage{multirow}
\usepackage{algorithm} 
\usepackage{algorithmic} 
\usepackage{mdframed} 
\usepackage{tcolorbox} 
\usepackage{enumitem}
\usepackage{hyperref}

\title{IDA: Breaking Barriers in No-code UI Automation Through Large Language Models and Human-Centric Design}

\author{
Segev Shlomov$^1$
\and
Avi Yaeli$^1$
\and
Sami Marreed$^1$
\and
Sivan Schwartz$^{1,2}$
\and
Netanel Eder$^1$
\and
Offer Akrabi$^1$
\and
Sergey Zeltyn$^1$\\
\affiliations
$^1$IBM Research\\
$^2$Technion – Israel Institute of Technology
\emails
segev.shlomov1@ibm.com,
aviy@il.ibm.com,
sami.marreed@ibm.com,
sivan.s@campus.technion.ac.il,
netanele@il.ibm.com,
offer.akrabi@il.ibm.com,
sergeyzil.ibm.com
}

\begin{document}

\maketitle

\begin{abstract}

Business users dedicate significant amounts of time to repetitive tasks within enterprise digital platforms, highlighting a critical need for automation.  Despite advancements in low-code tools for UI automation, their complexity remains a significant barrier to adoption among non-technical business users. However, recent advancements in large language models (LLMs) have created new opportunities to overcome this barrier by offering more powerful, yet simpler and more human-centric programming environments.
This paper presents IDA (Intelligent Digital Apprentice), a novel no-code Web UI automation tool designed specifically to empower business users with no technical background. IDA incorporates human-centric design principles, including guided programming by demonstration, semantic programming model, and teacher-student learning metaphor which is tailored to the skill set of business users. By leveraging LLMs, IDA overcomes some of the key technical barriers that have traditionally limited the possibility of no-code solutions.  We have developed a prototype of IDA and conducted a user study involving real world business users and enterprise applications. The promising results indicate that users could effectively utilize IDA
to create automation. The qualitative feedback indicates
that IDA is perceived as user-friendly and trustworthy.
This study contributes to unlocking the potential of AI assistants to enhance the productivity of business users through no-code user interface automation.

\end{abstract}

\maketitle

\section{Introduction}
\label{Introduction}


UI Automation emerges as a critical solution for business users overwhelmed with repetitive tasks across enterprise digital platforms. By automating repetitive work such as data entry, copying and pasting data, form filling, and report generation, businesses can focus on more strategic initiatives and higher-value tasks. Beyond just streamlining existing work, UI Automation is also key to unlocking the value of digital workers, and transforming how businesses operate.

However, the adoption of low-code automation tools by business users (i.e., non-technical personnel such as bankers, lawyers, and HR managers) remains limited due to the persistence of technical complexities that demand a programming mindset \cite{Rokis2022challenges,luo2021characteristics,sahay2020supporting}. Although recent no-code solutions aim to simplify web automation, they often oversimplify, failing to meet real-world application needs \cite{fischer2021diy,li2019pumice,pu2022semanticon}. Key hurdles for these users include defining stable UI selectors, crafting comprehensive specifications, translating presentation layers into business logic, and navigating the software development lifecycle.  Given recent advancements in LLMs, could they potentially overcome the technical challenges and simplify no-code tools to better suit the skills of non-technical users?

This paper introduces IDA (Intelligent Digital Apprentice), an innovative no-code UI automation tool designed for non-technical business users, facilitating the creation of web automation without prior programming knowledge. IDA leverages a human-centric design, drawing inspiration from the natural way business users instruct juniors, encapsulating this dynamic in a teacher-student relationship where the business user, referred to as "Cassie", acts as the teacher guiding IDA, the student. IDA incorporates principles of guided programming by demonstration, semantic programming model and multi-modal user interfaces, and aims to resonate with Cassie's technical abilities and instructional expectations.  

A prototype of IDA was developed embodying these principles, and novel algorithms powered by LLMs were designed and implemented to overcome the previously highlighted technical barriers.  A qualitative user study with real-world enterprise business users validated our hypothesis that IDA could support real-world use cases, demonstrating IDA’s ability to enable users to self-serve building automations effectively, and highlighting its intuitive nature and positive reception in terms of value, trust, and user experience. This paper main contributions are:
\begin{itemize}[leftmargin=*]
\setlength\itemsep{-0.1em}
\item Introduction of IDA's human-centric design, leveraging guided programming by demonstration (PbD), semantic programming, and a teacher-student metaphor, enhancing accessibility for non-technical users.
\item Development of the IDA prototype with innovative algorithms Semantic Element Understanding and guided Programming by Demonstration, addressing the technical challenges of no-code automation.
\item A user study with real business users, demonstrating IDA's effectiveness, ease of use, and trustworthiness in real-world scenarios, highlighting its potential to improve business productivity.
\end{itemize}

\section{Literature Review}
\label{Related Work}

Our work lies in the intersection of Human-computer interaction and UI automation. As the literature on these domains is vast, we identify gaps in the literature and draw our design inspirations from these areas.

\textbf{Web Automation} primarily involves the automatic control of web browsers to perform repetitive tasks such as data entry, content scraping, and interaction with web applications \cite{sugiura1998internet,little2007koala,leshed2008coscripter,le2014flashextract}. Early web automation tools were primarily developer-centric, requiring a good understanding of web technologies such as HTML, CSS, and JavaScript.
However, these tools pose challenges for non-technical users and it was even shown in \cite{krosnick2021understanding} that experienced programmers have difficulty in writing web automation using these frameworks, specifically in defining stable selector anchors for UI elements.

\textbf{Low-Code and No-Code}
(LCNC) platforms have transformed the digital labor domain, making it accessible for individuals to devise automation workflows with minimal coding. These platforms offer visual development environments (i.e., visual programming), allowing users to create applications through intuitive interfaces and pre-built elements, minimizing the need for extensive programming skills \cite{desmond2022no}. Low-code platforms such as Microsoft Power Automate \cite{MPO}, \citeauthor{automationanywhere},  \citeauthor{uipath}, and \citeauthor{zapier} enable users to develop automation workflows with basic coding, with a simple drag \& drop and configure experience, offering greater customization and flexibility. 

Despite their convenience, automations developed through these means may sometimes lack the sophistication and flexibility found in solutions coded by expert programmers. In addition, users still need to understand some basic programming concepts such as variables, constants, flow control and parameters which in practice means that non-technical business users are unable to create automation with them and still struggle with the nuances of complex business automation.

\textbf{Programming by demonstration} (PbD) refers to the technique where users can create programs by demonstrating the desired task instead of writing code \cite{cypher1993watch}. Users perform specific tasks, and the system generates the corresponding code or script automatically \cite{hirzel2022low,sereshkeh2020vasta}. For example, a user might demonstrate how to sort and organize data in a spreadsheet, and the system would generate a script that can perform this task automatically in the future. Over the years, tools like \citeauthor{uipath}, \citeauthor{imacro}, and  Vegemite adopted the PbD approach.
PbD is primarily utilized to simplify the process of programming, making it accessible to individuals who may not have extensive coding skills. 
\cite{dong2022webrobot} presents a formal foundation for reasoning about web RPA programs and proposes a program synthesis algorithm called speculative rewriting. 
\cite{pu2022semanticon} introduced SemanticOn, a system designed for semantically-driven web automation. Users initiate the process by specifying semantic conditions either through natural language prompts or by highlighting information on a webpage. Utilizing similarity-based computer vision and NLP techniques, these inputs are encoded into the program’s conditions. Users then demonstrate the desired actions on a target website with the aid of WebRobot, a low-code tool, creating automation programs without programming knowledge.

Recently, \cite{pu2023dilogic}, introduced DiLogic, a guided PbD system that semantically segments input data into manageable steps for user-friendly task tracking. Users demonstrate web interactions for each step, and the system identifies patterns in user demonstrations to generate automation programs, creating a catalog that maps task steps to UI actions for future reference. During automation, DiLogics matches each task step to this catalog, utilizing appropriate program logic based on semantic similarity.

However, typically in PbD, the generated code may not always be as efficient or clean as that written by an experienced programmer. This inefficiency could lead to performance issues, especially in large and complex automations. PbD also might not handle basic errors such as "element not found", which can result in less robust and reliable solutions. In addition, one of the common limitations of PbD is the ability to generalize demonstrations to broader use cases and support conditional or complex flows.
IDA, based on guided PbD, offers an all-encompassing pipeline for web automation, spanning from input data processing to customizing the program based on user demonstrations, merging multiple demonstrations, and refining and managing errors.

\textbf{LLM and LMM based Automation}
has revolutionized the way users interact with systems, enabling more human-like Interaction. \citeauthor{chatgpt,bard} and alike \cite{touvron2023llama} are examples of advanced language models that can interpret and generate human-like text, which can be harnessed for automation tasks. Autonomous agents, using LLM and LMM, allow users to describe high-level natural language intents in natural language, and the system decomposes the intent (planning) to a set of UI actions \cite{zheng2024gpt,wang2024mobile,he2024webvoyager,koh2024visualwebarena}.

Knowledgeable agents, such as \citeauthor{adept} also employ fine-tuned LMMs for downstream UI automation activity. These agents possess knowledge about what should be interacted with at every step of the automation and try to ground the instruction into the UI \cite{li2020mapping,wang2023enabling}.


However, autonomous agent approaches provide limited user control outside of describing the task in natural language and cannot be easily adapted to each specific user's needs and knowledgeable agents still struggle in grounding UI elements and thus lose trust quickly.

As trust is a critical factor in the adoption and continued use of automation tools, especially when these tools are meant to take over tasks traditionally executed by humans, ensuring transparency, reliability, providing clear feedback, and allowing users to have control are vital strategies. Research by \cite{schwartz2023enhancing} delves into that methodologies and best practices. We tried to follow their inspiration and build IDA in a trusted manner. Concluding, while significant advancements have been made in the field of automation and user-centric development tools, gaps remain, especially in creating tools that non-technical users find intuitive, efficient, and trustworthy. This paper aims to address some of these challenges by introducing IDA, a system designed with the business user at its core.

\section{Reference scenario}
\label{Reference Scenario}

This section outlines a reference scenario, HR Candidate Screening, to illustrate the system design and implementation. The scenario is modeled on typical \textbf{HR candidate screening processes} conducted by Talent Acquisition Specialists ("Cassie").
Cassie employs two main tools for screening: (1) Excel for tracking candidates and (2) OrangeHRM, an HR Management application, for detailed candidate information. Upon receiving a candidate list in Excel, she must decide on each candidate's next step by consulting their status in OrangeHRM.

\begin{figure}[ht!]
\centering
    \includegraphics[width=\linewidth]{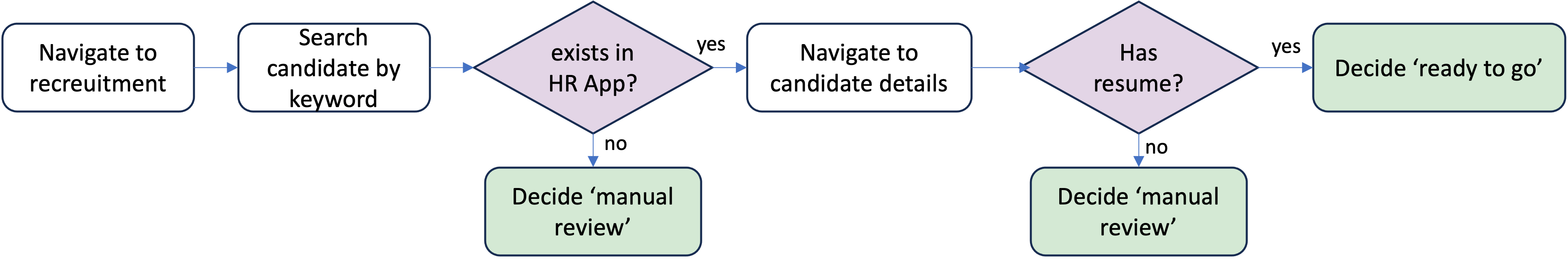}
    \caption{Decision-making flowchart in OrangeHRM: Actions (white), observed states (purple), and decision outcomes (green).}
    \label{fig:HR Screening - Flow}
\end{figure}

\begin{figure}[ht!]
\centering
    \includegraphics[width=\linewidth]{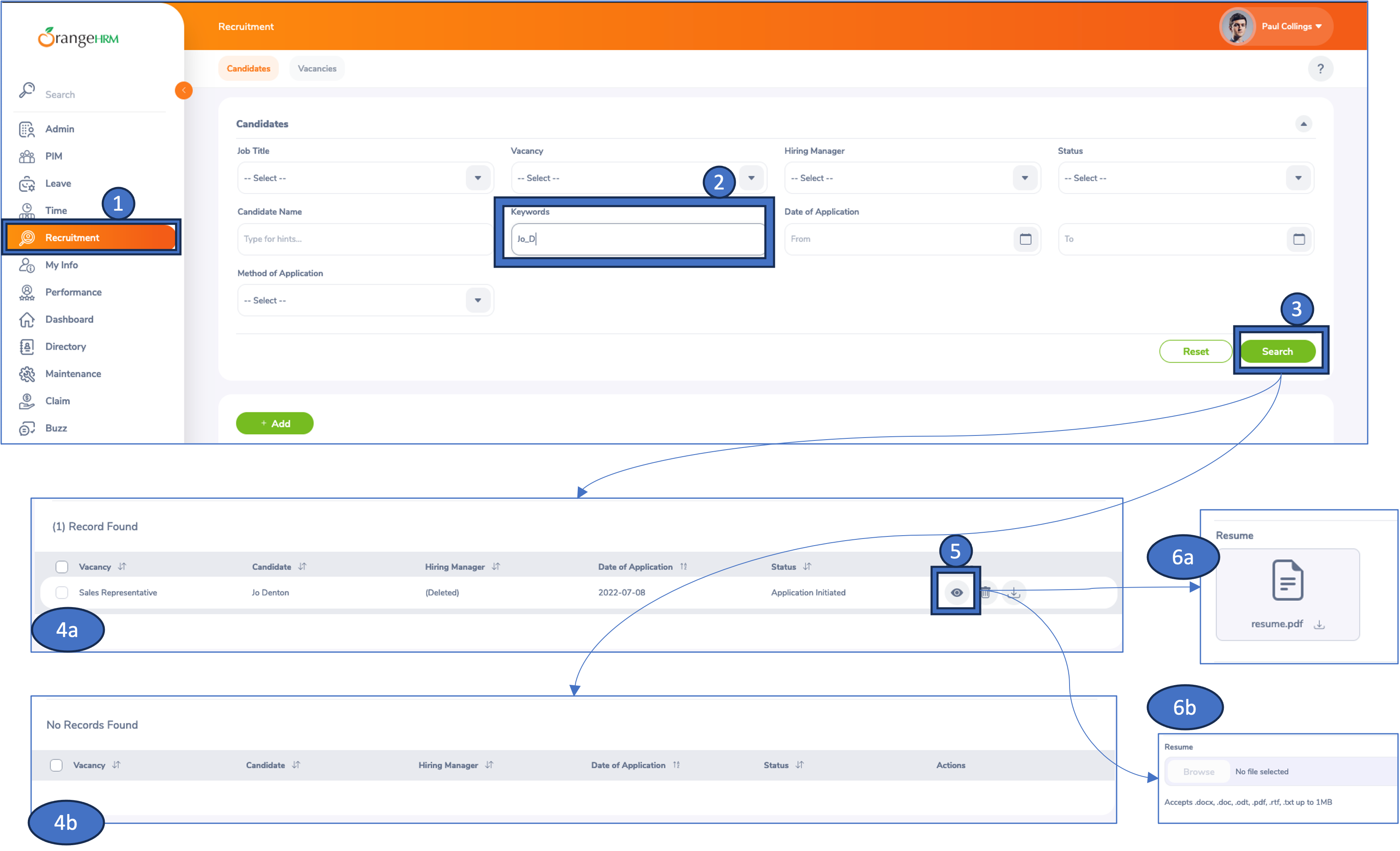}
    \caption{UI interaction steps in OrangeHRM for candidate decision-making. Includes candidate lookup, search results, and resume status checks.}
    \label{fig:HR Screening - UI}
\end{figure}

The decision-making process involves checking whether candidates have submitted their resumes through OrangeHRM, as depicted in Figure \ref{fig:HR Screening - Flow}. This involves Cassie performing a name-based search in OrangeHRM, examining search results, and, if a candidate exists, reviewing their detailed profile for resume submission status. This information is then used to update each candidate's status in the Excel file, categorizing them for either immediate proceeding (\textit{ready to go}) or \textit{manual review}. Figures \ref{fig:HR Screening - Flow} and \ref{fig:HR Screening - UI} visualize this workflow, including UI elements Cassie interacts with, such as the search function and candidate details, to make informed decisions.

\section{Design principles and requirements}
\label{Design principles and requirements}


This section outlines the design principles and requirements of IDA, emphasizing design goals (DGs) that address the challenges faced by non-developers with previous no-code solutions. Recognizing the technical complexity that hindered adoption, IDA is crafted around the natural capabilities and technical proficiency of users like Cassie. Drawing inspiration from how business users naturally teach and learn tasks on digital platforms, we aimed to create a user experience that feels intuitive and easy for users to embrace.

\subsection{Guided programming by demonstration}

Previous work \cite{dong2022webrobot,li2019pumice} found that while business users can illustrate specific situations through individual demonstrations, generalizing these into complete program specifications still poses challenges.  Further drawing from \cite{harel2003specifying} insights on predefined inputs and outputs facilitating complete program behavior specification, IDA incorporates a guided PbD approach, leading to these design goals:
\\ \noindent \textbf{DG1: Multiple Scenarios}: Support recording various user demonstrations for different business contexts, synthesizing them into a cohesive program.
\\ \noindent \textbf{DG2: Conditional Branching}: Enable demonstrations across diverse scenarios with conditional logic based on UI elements or input values, including system assistance for aligning and branching steps.
\\ \noindent \textbf{DG3: Scenario Completion and Conflict Resolution} Recommend more demonstrations to cover all input possibilities and conditional paths and identify conflicting instructions.
\\ \noindent \textbf{DG4: Lifecycle Guidance}: Offer non-technical users step-by-step guidance through the development lifecycle to ensure clarity at every stage.

 
\subsection{Semantic programming model}
This subsection introduces IDA's semantic programming model, aiming to bridge the technical skill gap in UI element selection and complex expression definition. It assumes that users would benefit from referring to UI elements by semantic names and roles, leading to these design goals:


\noindent\textbf{DG5: Semantic Element Identification}: Automatically identify and name UI elements based on visible text, linking each to its type and semantic role.
Further addressing complex variable and expression challenges, the goals expand to:
\\ \textbf{DG6: Parameter Mapping}: Automate the mapping of demonstrated input values to named parameters within the automation.\\
\textbf{DG7: Abstract Representation}: Offer a higher-level action model that simplifies complexity, aligning closely with the user's mental model of demonstrations and conditions.


Addressing the challenge of conditional execution based on visible elements, such as differentiating conditions based on search results, IDA aims to semantically understand and guide users in defining conditions based on visual context:
\\\textbf{DG8: Semantic Guidance}: Provide in-context interactive guidance for identifying and setting conditions on semantic objects, simplifying the process without requiring users to handle complex data scraping or expressions.


\subsection{Teacher student metaphor with feedback loops}

We explore the implementation of a teacher-student metaphor in IDA, drawing inspiration from the interactive training and validation methods business users employ for effective learning. Our goal is to foster trust and ensure the accuracy of IDA's learning process through well-integrated feedback loops. The design goals to support this include:
\\\noindent\textbf{DG9: Learning Confirmation}: Implement live feedback through visual and auditory cues to assure Cassie of the correct learning of demonstrated steps by IDA.
\\\noindent\textbf{DG10: Intuitive Visualizations}: visual representations of both the recorded scenarios and the synthesized program, allowing Cassie to easily verify IDA's learning progress.
\\\noindent\textbf{DG11: Demonstration Playback}: Enable IDA to replay its execution of automation tasks using sample data, facilitating Cassie's validation of the automation accuracy.

\textbf{Multi-modal interfaces:}
IDA incorporates voice alongside text as input and output options, reflecting the natural use of language in business for communication and commands. It offers both visual and auditory feedback for a versatile and accessible user experience, which, although not its main focus, has been well-received by users.

\section{IDA}
\label{Implementation}


This section details IDA's functionality and its application in automating a candidate screening scenario (see Section \ref{Reference Scenario}). We focus on demonstrating IDA's user experience, through Cassie's interactions, and discuss key underlying algorithms that facilitate this process. The notation [DGx] references the design goals addressed herein. A demonstration video of IDA is available for further insight at \textbf{\href{http://tinyurl.com/ida-demo}{tinyurl.com/ida-demo}}.


IDA is built around three core components: a browser JavaScript extension, a React-based client UI, and a Python-powered back-end server. This architecture is designed to simulate an apprentice-like experience for users, with IDA effectively acting as an assistant that provides immediate feedback during the automation process.
The Browser Extension observes and relays user interactions with the web application to the server.  The 
Client UI serves as the control panel for users, guiding them through the automation's lifecycle stages—Define, Teach, Validate—with visual cues to indicate IDA's learning status.
The Back-End Server functions as the system's brain, processing interactions, managing UI control, executing business logic, and storing automation data.

\begin{figure}[ht!]
    \includegraphics[width=\linewidth] 
    {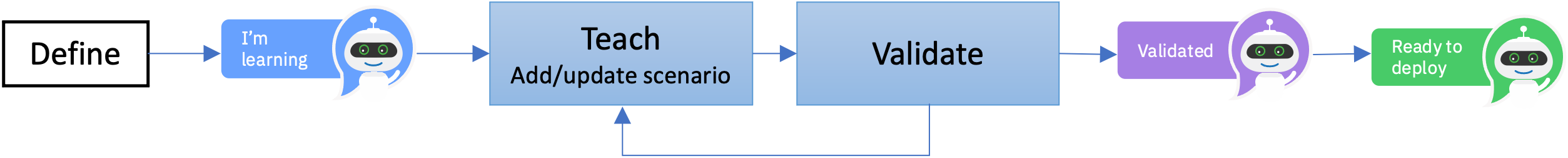}
    \caption{\textbf{IDA's Automation Lifecycle}: Illustrates IDA's step-by-step guidance from defining ('I'm Learning' state) to validation and deployment readiness ('Ready to Deploy' state), ensuring a user-validated development process.}
    \label{fig:ida_lifecycle}
\end{figure}

The client UI is strategically placed adjacent to the HR application, enhancing the user experience by allowing seamless interaction without the need to toggle between windows. 
IDA guides users through the Define-Teach-Validate stages [DG4] of automation creation, offering visual indicators of its learning state, such as an "I'm learning" icon (Figure \ref{fig:ida_lifecycle}). This iterative process culminates in the automation being marked "Ready to deploy" once validated by the user and IDA.

\begin{figure}[ht!]
\centering
    \includegraphics[width=\linewidth]{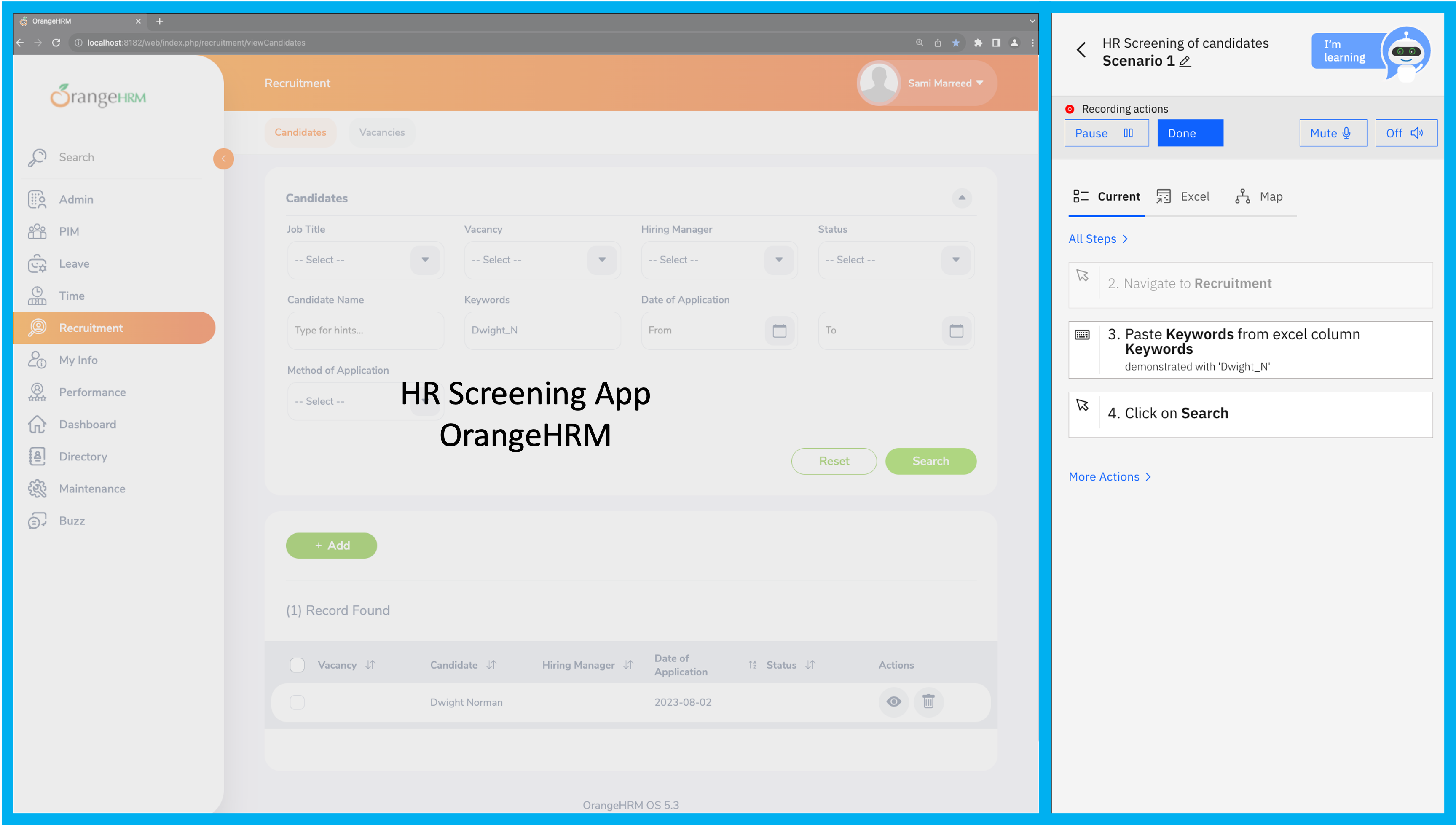}
    \caption{\textbf{IDA Integration for Seamless Workflow}: Showcases the side-by-side setup of the IDA client and HR application, enabling Cassie to interact with both simultaneously for a seamless demonstration and feedback experience.}
    \label{fig:ida_system}
\end{figure}

\subsection{IDA User Experience}

\subsubsection{Define}
Cassie starts by naming and describing the automation, then selects its type (e.g., "Excel $\rightarrow$ Automation $\rightarrow$ Excel") utilizing a template for iterating over Excel rows. She uploads a sample Excel file followed by the possible decision values, and moves IDA to advance to the "Teach" stage.  Further details are provided in the supplementary materials.

\subsubsection{Teach}  
Cassie demonstrates the process by teaching IDA various scenarios and decisions, like "ready-to-go" and "manual-review". IDA's in-context guidance helps Cassie define actions and conditions directly on the UI without coding, facilitating an intuitive teaching process. She records steps for each scenario, with IDA providing feedback and suggesting additional scenarios to ensure comprehensive coverage.

During teaching, Cassie teaches IDA how to perform the automation based on different business situations (aka scenarios hereunder). In the case of HR Candidate Screening, Cassie must provide multiple demonstrations in order to cover all the scenarios of the process.  Cassie instructs IDA on two pathways to the manual-review decision: (1) for a candidate without a resume (\textit{user exists $\rightarrow$ no resume $\rightarrow$ manual-review}), and (2) for a candidate not found in the system (\textit{user doesn't exist $\rightarrow$ manual-review}), labeled as manual-review1 and manual-review2, respectively.


\begin{figure}
    \centering
    \includegraphics[width=\linewidth]{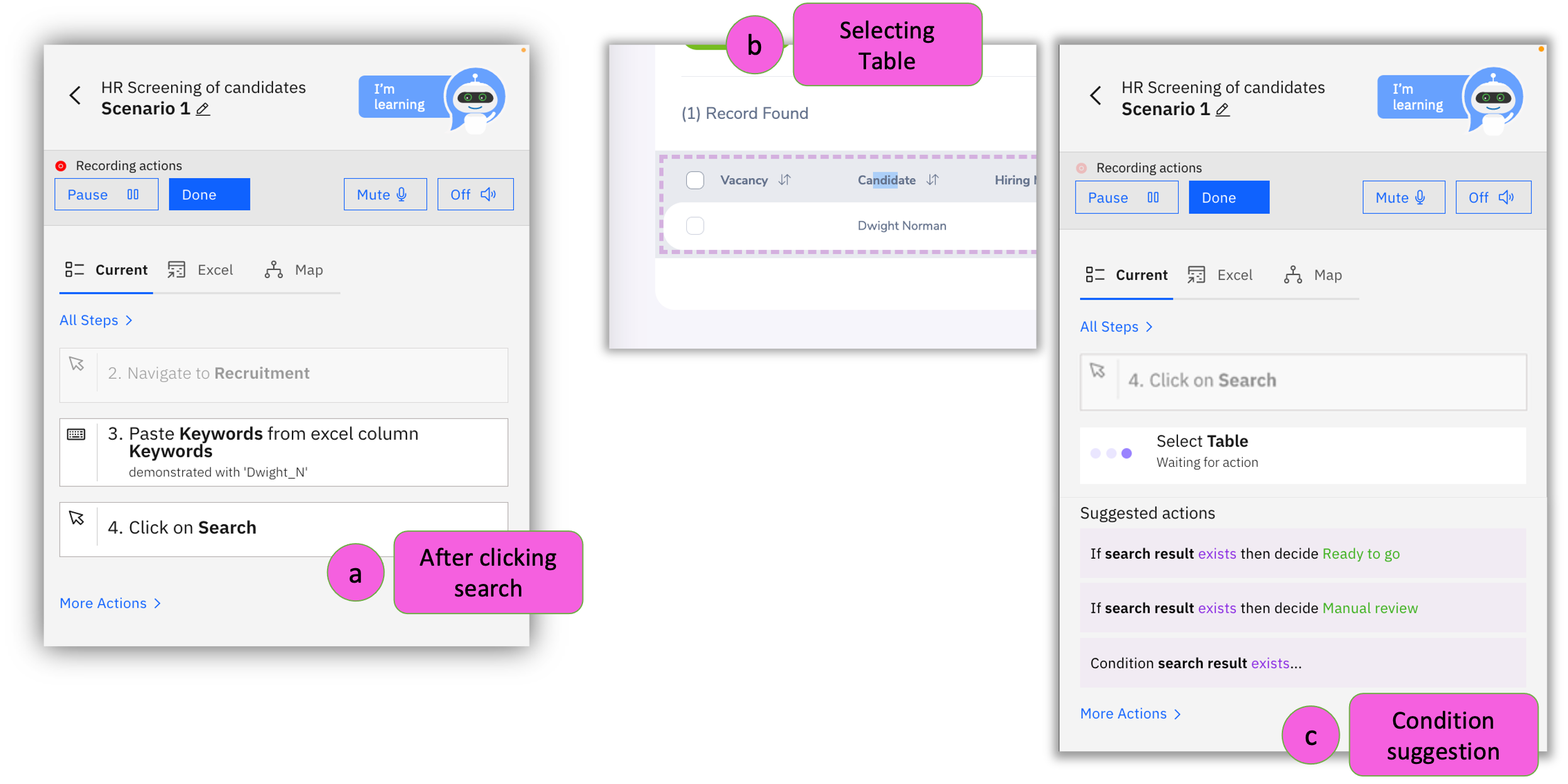}    
    \caption{\textbf{IDA's in-context Semantic Guidance}: In the manual-review1 scenario, IDA (a) displays captured steps, (b) Cassie identifies the search table by selecting the header, and (c) IDA suggests actions for Cassie to choose verbally or by click. Semantic conditions are highlighted in purple, decisions in green.}
    \label{fig:teach_in_context}
\end{figure}

\textbf{In-context Scenario Teaching.} Cassie demonstrates the manual-review scenario by recording her actions to navigate to the search candidate form. She initiates by clicking the record button, then proceeds to the Recruitment tab and inputs a keyword from a sample Excel sheet into the search field to locate a candidate. Given IDA's understanding of the Excel data schema, it interprets the input value as an external parameter for the automation [DG6]. Throughout the process, IDA offers immediate feedback for each step, allowing Cassie to verify its accuracy and comprehension [DG9-10, DG12-13]. IDA also enhances understanding by linking web elements to their labels, presenting Cassie with intuitive names for the steps recorded [DG5].

As Cassie reviews the steps, she prepares to teach IDA conditional branching based on search results. Recognizing tables and other complex objects, IDA simplifies interaction by identifying these elements and enabling Cassie to operate at a semantic level without coding [DG4, DG8]. To define a condition, she selects the candidate column in the search result.

As shown in Figure \ref{fig:teach_in_context}, IDA then visualizes the selected table element and its context, offering Cassie actionable suggestions like "Condition search results exist" for further scenario development [DG2]. This capability stems from IDA's ability to detect the semantic element, its function, and state, recommending relevant actions. Following this, Cassie demonstrates steps leading to the resume attachment field, with IDA recognizing the attachment input and its state based on Cassie's input, streamlining the teaching process.

\textbf{Guiding Cassie through Multiple Scenarios.} After Cassie finishes the manual-review1 scenario, IDA summarizes the steps taken and suggests additional scenarios for demonstration. These recommendations cover unexplored areas in the decision-making process and conditional states of semantic objects [DG1, DG3]. IDA prompts Cassie to demonstrate the "ready-to-go" scenario defined earlier. It suggests the "search results exist $\rightarrow$ Resume NotFound" scenario, based on an analysis of previously demonstrated elements and conditions. This approach acts as a reminder for Cassie to begin the second scenario, emphasizing the importance of consistency with earlier demonstrations. For instance, in manual-review2, she should replicate the initial steps until reaching the shared branching point of the first scenario, then proceed to document steps when no search results are found. Similarly, for the "ready-to-go" scenario, she must follow through to the decision-making step regarding the resume element.  Should IDA detect any discrepancies with the established flow, it flags the scenario with an error, ensuring accuracy and consistency across demonstrations.

\textbf{Visualizing the automation program.} After Cassie records multiple scenarios, IDA synthesizes these into a semantic program and showcases an automation map. This visualization helps Cassie review the automation's behavior across different scenarios [DG7]. The map illustrates key decision points, such as a "Ready to go" when a "resume" exists and a "manual review" decision when it does not, offering Cassie a clear view of the automation's logic and flow.  See supplementary material for further details.


\subsubsection{Validate}
\begin{figure}[h!]
    \centering
    \includegraphics[width=\linewidth]{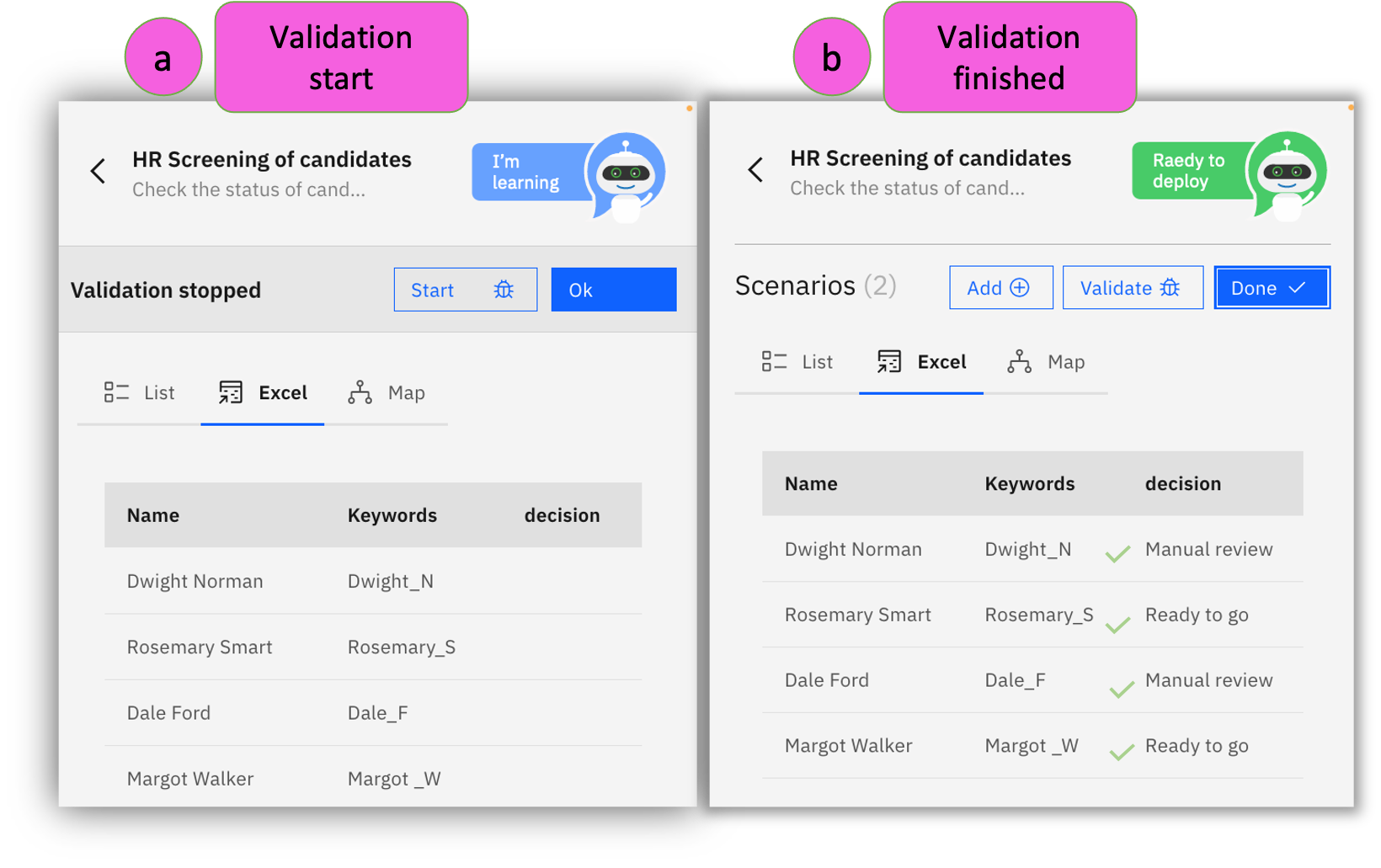}
    \caption{\textbf{Validation}: Initially, IDA is in the "I'm learning" state (a). Once Cassie starts the validation, IDA methodically demonstrates the automation across all input examples, updating each decision. Upon completion, IDA shifts to "Ready to deploy" (b), indicating its readiness after thorough validation.}
    \label{fig:validation}
\end{figure}



After teaching, Cassie initiates validation by clicking the 'validate' button (Figure \ref{fig:validation}), ensuring that the application's state matches the automation's start point. IDA interacts with the UI step-by-step, allowing Cassie to monitor the progress. For Excel-based automation, progress is displayed both in the Excel tab and on the automation map, highlighting the column IDA fills and the UI step being validated [DG11].
 
\subsection{IDA Algorithms}

\subsubsection{Semantic Element Understanding}

When Cassie interacts with an application UI, IDA must interpret her actions accurately and display their semantic representation in the scenario list view.  For basic UI elements like inputs, buttons, and labels, IDA employs CSS selectors for the DOM and Shadow-DOM (which shields components from direct querying of the DOM). This method relies on detailed mappings of HTML standards and popular web frameworks [DG5]. Moreover, IDA uses pixel-based rules for aligning UI components with their semantic descriptors, such as linking an input box to its label, enhancing the identification and association process.

\begin{figure}[h!]
    \centering
    \includegraphics[width=\linewidth]{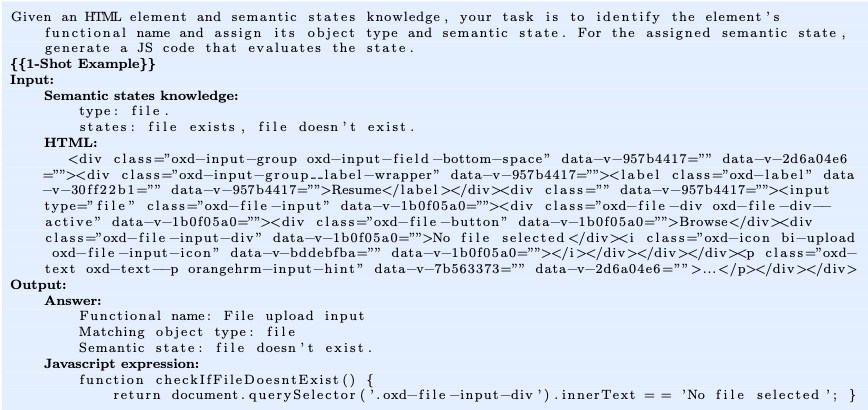}
    \caption{The prompt of semantic objects and states detection. The output was generated with Llama2 70B.}
    \label{llm-p}
\end{figure}

For in-context guidance with complex semantic UI elements such as search result tables or file attachment indicators, IDA leverages a novel algorithm powered by a LLM. This algorithm enables IDA to recognize these elements, ascertain their names, and interpret their states within the context of the current application and user demonstration [DG8]. It does so by defining a set of semantic objects and their potential states (for example, a search result might be in a "one record" or "multiple records" state, while a file attachment might be "present" or "absent"). Utilizing a LLM, IDA parses HTML snippets to identify semantic objects and their states, assigns user-friendly names to these objects, and generates JavaScript code capable of assessing the object's state in real-time. This process is depicted in Figure \ref{llm-p}, where the LLM identifies the semantic element type (e.g., "table"), assigns a user-friendly name ("Companies table"), and determines its semantic state ("has multiple records").

This semantic programming model marks a paradigm shift from the traditional, highly technical approach to UI automation, where users needed a deep understanding of HTML syntax and coding. Previously, creating automation like checking a file's state required intricate coding, resulting in complex and fragile scripts.  With IDA, however, the process is simplified to describing the desired outcome ("what") while the underlying LLM-based algorithm handles the complexities of implementation ("how"). This breakthrough not only mitigates the barriers to adoption by non-technical business users but also enhances the resilience and reusability of automation scripts, significantly improving the return on investment and user satisfaction with automation technologies.

\subsubsection{Program synthesis from multiple demonstrations}

When Cassie provides multiple demonstrations, IDA synthesizes them into a unified program, identifying and alerting inconsistencies. The synthesis algorithm merges each new scenario with the existing program, aligning steps with common prefixes until a divergence is encountered. Minor inconsistencies are tolerated. However, significant discrepancies, such as differing initial actions, flag the program as inconsistent, prompting Cassie for review or re-recording.

The algorithm also adeptly handles conditional statements, merging scenarios with conditions like the presence or absence of search results into a single step with dual branches. This is achieved through label matching for semantic object identification, ensuring that even complex conditional logic is integrated smoothly.
Through iterative processing of each scenario, IDA creates a comprehensive executable program that encapsulates the logic of all provided demonstrations, ensuring a robust and cohesive automation solution.

\section{User study evaluation}
\label{User Study Evaluation}

The objective is to evaluate IDA's effectiveness and user experience for non-technical business users in automating tasks of varying complexity. It is structured around two primary objectives:
\\ \noindent\textbf{Objective 1:} Assess the capability of non-technical business users to successfully create automation with IDA, demonstrating that the system's no-code interface is intuitively designed to match their skill set.
\\ \noindent\textbf{Objective 2:} Collect user feedback on IDA's overall value and user experience, measuring satisfaction and trust.

\subsection{Methodology}
We recruited eight business users, all [ANONYMOUS] employees, who volunteered to participate in the study. The group consisted of five females and three males, with an age range of 25 to 57 years (mean age: 39.1, SD=10.2). Their educational backgrounds varied, including three with bachelor’s degrees, four with master’s degrees, and one bachelor student. These participants, selected for their roles in Marketing, Legal \& IP, and Administration, had no prior programming or automation training, making them ideal for assessing the intuitiveness of the IDA no-code platform. 

\textbf{Procedure}
The evaluation consisted of a concise session, approximately 60 minutes long, accommodating the busy schedules of business users in an  enterprise. Prior to the session, participants completed a questionnaire to provide demographic information, technical background, and informed consent (see supplementary material for more details).  Each session was structured to assess both the practical use of IDA and participant feedback:

\textbf{Training and Evaluation}.
Participants underwent a brief 10-minute training session on IDA to familiarize themselves with its no-code interface and basic functionalities. Afterward, they were tasked with creating an automation independently to evaluate their ability to apply what they learned. We designed two tasks aimed to mirror real-world automation complexity without requiring programming expertise:\\  
\textbf{Medium complexity task:} This task involved developing an automation that reads names of cities from Excel, lookup these cities on the weather.com website, scrapes the current temperature, and saves it to a new column in Excel.\\  
\noindent\textbf{High complexity task:} This task involved developing an automation for the reference HR Screening scenario described in Section \ref{Reference Scenario}.  This included reading candidates' names from Excel, searching them in the HR application, navigating to their details page, evaluating if they already provided a resume, and conditionally on this evaluation, updating the corresponding decision back in Excel.  

To evaluate objective 1, we defined a success metric on a binary scale, where a score of 1 indicates successful completion of a task, and a score of 0 indicates failure.  

\textbf{Qualitative Feedback}.
After the practical tasks, participants completed a qualitative questionnaire to provide insights into their user experience, including IDA's intuitiveness, satisfaction, and perceived value (See supplementary material for the full questionnaire).   

\textbf{Data collection} focused on task completion time, success rates in the evaluation tasks, and responses from the pre- and post-session questionnaires. The post-session questionnaire included both Likert scale items and open-ended questions to capture comprehensive user feedback.


\subsection{Results}

\subsubsection{Task Success}
Table \ref{tbl1} presents the success and duration for each participant per task, and Table \ref{tbl2} presents descriptive statistics for all participants.  The results show that all participants were able to complete the medium-complexity task successfully, and only one participant did not succeed in completing the high-complexity task.  These results support IDA's effectiveness and intuitive design for non-technical business users.

\begin{table}[h!]
\caption{Success level and duration to complete each task per subject.}
\label{tbl1}
\resizebox{\linewidth}{!}{
\begin{tabular}{ccccc} \hline
           & \multicolumn{2}{c}{\textbf{First Task - Medium Complexity}} & \multicolumn{2}{c}{\textbf{Second Task - High Complexity}} \\ 
Subject ID & Success Level           & Duration (min)           & Success Level           & Duration (min)           \\\hline
1          & 1                       & 2.57                     & 1                       & 10.15                    \\
2          & 1                       & 14.62                    & 0                       & 9                        \\
3          & 1                       & 5.6                      & 1                       & 3.77                     \\
4          & 1                       & 6.68                     & 1                       & 4.42                     \\
5          & 1                       & 2.83                     & 1                       & 3.87                     \\
6          & 1                       & 2.92                     & 1                       & 9.12                     \\
7          & 1                       & 7.62                     & 0                       & 8.5                      \\
8          & 1                       & 6.38                     & 1                       & 8                       \\\hline 

\end{tabular}}
\end{table}


\begin{table}[h!]
\caption{Descriptive statistics for success and duration per task.  }
\label{tbl2}
\resizebox{\linewidth}{!}{

\begin{tabular}{llcccccc}
\hline
\textbf{Measure} & \textbf{Task type} & \textbf{N} & \textbf{Mean} & \textbf{SD} & \textbf{Median} & \textbf{Min} & \textbf{Max} \\ \hline
\multirow{2}{*}{\textbf{Success level}} & Medium & 8 & 1 & 0 & 1 & 1 & 1 \\
 & High & 7 & 0.85 & 0.35 & 1 & 0 & 1 \\ 
 
 \hline
 
\multirow{2}{*}{\textbf{Duration (min)}} & Medium & 8 & 6.15 & 3.94 & 5.99 & 2.57 & 14.62 \\
 & High & 7 & 6.83 & 2.72 & 8.00 & 3.77 & 10.15 \\ \hline
\end{tabular}}
\end{table}

\subsubsection{User Feedback}
Post-evaluation, participants provided feedback through a questionnaire assessing IDA's usability and their experience. Specific feedback and scores for each statement are detailed in Table \ref{tbl3}.  The feedback was overwhelmingly positive, with participants finding IDA comfortable and easy to use, as indicated by high scores on questions related to system usability, ease of learning, and the minimal cognitive effort required (S1, S5, S7, S8, S11). The graphical design was easily understood and used appropriately (S1, S2), and the use of sound was found to be helpful (S6, S10). Participants expressed a strong willingness to incorporate IDA into their daily workflows (S9), highlighting its potential value in automating routine business tasks. Nevertheless, we did not observe positive significant results for the ability of IDA to recover from mistakes, an issue that we will address in the future. 

At the end of the questionnaire, we asked qualitative questions of the users, among them, we asked whether the user would trust our system in his daily routine as a business user, and all of the subjects agreed. When we further asked for what kind of business goals the subject will use our system the report included: scheduling repetitive meetings, excel budget updates based on workers' reports, excel data filtering, payment for patent licensing on an official web page, and report employee work hours. 
Given this qualitative feedback, we are positive that IDA will contribute to business users in many tasks and business users will trust it after appropriate and short training as we conducted in our experiment. 

\begin{table}
\caption{Descriptive statistic of statements in a 5-point Likert scale (5 represents strongly agree).}
\label{tbl3}
\resizebox{\linewidth}{!}{
\begin{tabular}{llccccc}
\hline
\textbf{Statement} & \textbf{Question description} & \textbf{Average} & \textbf{SD} & \textbf{Median} & \textbf{Min} & \textbf{Max}  \\ \hline
\textbf{S1} & \textbf{\begin{tabular}[c]{@{}l@{}}I think the system was comfortable\\  and easy to use.\end{tabular}} & 4.25 & 0.46 & 4 & 4 & 5  \\
\textbf{S2} & \textbf{\begin{tabular}[c]{@{}l@{}}It was easy for me to understand \\ where each component I needed \\ was located on the screen.\end{tabular}} & 4.25 & 0.46 & 4 & 4 & 5 \\
\textbf{S3} & \textbf{\begin{tabular}[c]{@{}l@{}}I think that relying on technical \\ support for using a system like\\  IDA was necessary.\end{tabular}} & 2.63 & 1.06 & 2 & 2 & 5 \\
\textbf{S4} & \textbf{\begin{tabular}[c]{@{}l@{}}If I mismarked something, it \\ was easy for me to correct the\\  error and continue with the task.\end{tabular}} & 3.88 & 0.99 & 4 & 2 & 5 \\
\textbf{S5} & \textbf{\begin{tabular}[c]{@{}l@{}}I feel that performing the tasks\\ required a lot of effort from me.\end{tabular}} & 1.88 & 0.35 & 2 & 1 & 2  \\
\textbf{S6} & \textbf{\begin{tabular}[c]{@{}l@{}}Using sound helped me in \\ performing the task.\end{tabular}} & 4.25 & 0.71 & 4 & 3 & 5  \\
\textbf{S7} & \textbf{\begin{tabular}[c]{@{}l@{}}I think the system was too \\ complex to use.\end{tabular}} & 1.63 & 0.52 & 2 & 1 & 2  \\
\textbf{S8} & \textbf{\begin{tabular}[c]{@{}l@{}}I think I can use my own \\ abilities in such a system.\end{tabular}} & 4.38 & 0.52 & 4 & 4 & 5  \\
\textbf{S9} & \textbf{\begin{tabular}[c]{@{}l@{}}If I had access, I would use IDA\\ for automating my daily\\ workflow at work.\end{tabular}} & 4.63 & 0.52 & 5 & 4 & 5  \\
\textbf{S10} & \textbf{\begin{tabular}[c]{@{}l@{}}The use of sound did not contribute\\ to my task performance.\end{tabular}} & 1.50 & 0.76 & 1 & 1 & 3  \\
\textbf{S11} & \textbf{\begin{tabular}[c]{@{}l@{}}I feel that task performance \\ required a small allocation\\ of my attention resources.\end{tabular}} & 2.63 & 1.06 & 3 & 1 & 4  \\ \hline
\end{tabular}}
\end{table}

\subsubsection{Key Findings}

The evaluation showcases IDA's potential to be quickly learned and effectively used by business users, even within a limited timeframe. The positive qualitative feedback from expert users underscores the system's user-friendly design and the positive user experience it offers.  Participants' willingness to use IDA for their daily work tasks suggests a strong perceived value and trust in the system.  This underscores the practical relevance of IDA for business users, affirming its usability and potential for enhancing business processes without the need for programming expertise.

\section{Discussion}

In our exploration of no-code automation, IDA represents a significant step towards dismantling the barriers faced by non-technical business users in automating web UI tasks. While we have begun to unlock the potential for end-users to navigate the complexities of digital platforms without extensive coding knowledge, our work reveals an evolving landscape of challenges that must be addressed to fully realize this vision. The integration of PbD, a semantic programming model, and the utilization of LLMs has enabled IDA to adeptly handle common tasks such as clicking, typing, and semantic conditions. However, extending this capability to encompass a broader array of actions, manage multiple decisions and outcomes, and support complex compositions remains a work in progress. These enhancements are crucial for IDA to generalize effectively to real-world applications, where the diversity and complexity of user needs are vast.

To further refine IDA's capabilities, several improvements are discussed. A skeleton-first approach, allowing users to draft a high-level outline of their automation tasks in natural language, promises a structured and efficient process for creating automations. This approach will enable users to conceptualize and refine their automation strategies with greater clarity and purpose. Furthermore, the development of smarter merging algorithms that can discern user intentions and understand semantic contexts is essential for IDA to accurately generalize from multiple demonstrations, thereby enriching the user experience with a more intuitive automation process. Additionally, introducing a chat or dialog interface that supports multi-modal interactions could significantly enhance user engagement, offering a novel and accessible way for users to interact with and refine their automations.

Looking beyond the web UI domain, the ambition to extend IDA's functionality to other platforms, including desktop applications and mobile platforms,  presents an exciting frontier. Achieving this expansion will require a shift in focus from the predominantly DOM-based strategies currently employed to innovative approaches that  leverage computer vision and further integration of large multi-modal models. Such a transition would enable IDA to interpret and interact with screen content in a manner akin to human perception, broadening its applicability and marking a pivotal step towards a domain-agnostic automation model.

\bibliographystyle{named}
\bibliography{references}

\begin{thebibliography}{}

\bibitem[\protect\citeauthoryear{Adept}{2024}]{adept}
Adept.
\newblock \url{https://www.adept.ai/}, 2024.

\bibitem[\protect\citeauthoryear{{Automation Anywhere}}{Accessed October 2023}]{automationanywhere}
{Automation Anywhere}.
\newblock \url{https://www.automationanywhere.com/}, Accessed: October, 2023.

\bibitem[\protect\citeauthoryear{Bard}{Accessed October 2023}]{bard}
Google. Bard.
\newblock \url{https://bard.google.com/}, Accessed: October, 2023.

\bibitem[\protect\citeauthoryear{Cypher and Halbert}{1993}]{cypher1993watch}
Allen Cypher and Daniel~Conrad Halbert.
\newblock {\em Watch what I do: programming by demonstration}.
\newblock MIT press, 1993.

\bibitem[\protect\citeauthoryear{Desmond \bgroup \em et al.\egroup }{2022}]{desmond2022no}
Michael Desmond, Evelyn Duesterwald, Vatche Isahagian, and Vinod Muthusamy.
\newblock A no-code low-code paradigm for authoring business automations using natural language.
\newblock {\em arXiv preprint arXiv:2207.10648}, 2022.

\bibitem[\protect\citeauthoryear{Dong \bgroup \em et al.\egroup }{2022}]{dong2022webrobot}
Rui Dong, Zhicheng Huang, Ian~Iong Lam, Yan Chen, and Xinyu Wang.
\newblock Webrobot: web robotic process automation using interactive programming-by-demonstration.
\newblock In {\em Proceedings of the 43rd ACM SIGPLAN International Conference on Programming Language Design and Implementation}, pages 152--167, 2022.

\bibitem[\protect\citeauthoryear{Fischer \bgroup \em et al.\egroup }{2021}]{fischer2021diy}
Michael~H Fischer, Giovanni Campagna, Euirim Choi, and Monica~S Lam.
\newblock Diy assistant: a multi-modal end-user programmable virtual assistant.
\newblock In {\em Proceedings of the 42nd ACM SIGPLAN International Conference on Programming Language Design and Implementation}, pages 312--327, 2021.

\bibitem[\protect\citeauthoryear{Guilmette}{2020}]{MPO}
Aaron Guilmette.
\newblock {\em Workflow Automation with Microsoft Power Automate: Achieve digital transformation through business automation with minimal coding}.
\newblock Packt Publishing Ltd, 2020.

\bibitem[\protect\citeauthoryear{Harel and Marelly}{2003}]{harel2003specifying}
David Harel and Rami Marelly.
\newblock Specifying and executing behavioral requirements: the play-in/play-out approach.
\newblock {\em Software \& Systems Modeling}, 2:82--107, 2003.

\bibitem[\protect\citeauthoryear{He \bgroup \em et al.\egroup }{2024}]{he2024webvoyager}
Hongliang He, Wenlin Yao, Kaixin Ma, Wenhao Yu, Yong Dai, Hongming Zhang, Zhenzhong Lan, and Dong Yu.
\newblock Webvoyager: Building an end-to-end web agent with large multimodal models.
\newblock {\em arXiv preprint arXiv:2401.13919}, 2024.

\bibitem[\protect\citeauthoryear{Hirzel}{2022}]{hirzel2022low}
Martin Hirzel.
\newblock Low-code programming models.
\newblock {\em arXiv preprint arXiv:2205.02282}, 2022.

\bibitem[\protect\citeauthoryear{iMacro}{Accessed October 2023}]{imacro}
iMacro.
\newblock \url{https://www.progress.com/imacros}, Accessed: October, 2023.

\bibitem[\protect\citeauthoryear{Koh \bgroup \em et al.\egroup }{2024}]{koh2024visualwebarena}
Jing~Yu Koh, Robert Lo, Lawrence Jang, Vikram Duvvur, Ming~Chong Lim, Po-Yu Huang, Graham Neubig, Shuyan Zhou, Ruslan Salakhutdinov, and Daniel Fried.
\newblock Visualwebarena: Evaluating multimodal agents on realistic visual web tasks.
\newblock {\em arXiv preprint arXiv:2401.13649}, 2024.

\bibitem[\protect\citeauthoryear{Krosnick and Oney}{2021}]{krosnick2021understanding}
Rebecca Krosnick and Steve Oney.
\newblock Understanding the challenges and needs of programmers writing web automation scripts.
\newblock In {\em 2021 IEEE Symposium on Visual Languages and Human-Centric Computing (VL/HCC)}, pages 1--9. IEEE, 2021.

\bibitem[\protect\citeauthoryear{Le and Gulwani}{2014}]{le2014flashextract}
Vu~Le and Sumit Gulwani.
\newblock Flashextract: A framework for data extraction by examples.
\newblock In {\em Proceedings of the 35th ACM SIGPLAN Conference on Programming Language Design and Implementation}, pages 542--553, 2014.

\bibitem[\protect\citeauthoryear{Leshed \bgroup \em et al.\egroup }{2008}]{leshed2008coscripter}
Gilly Leshed, Eben~M Haber, Tara Matthews, and Tessa Lau.
\newblock Coscripter: automating \& sharing how-to knowledge in the enterprise.
\newblock In {\em Proceedings of the SIGCHI Conference on Human Factors in Computing Systems}, pages 1719--1728, 2008.

\bibitem[\protect\citeauthoryear{Li \bgroup \em et al.\egroup }{2019}]{li2019pumice}
Toby Jia-Jun Li, Marissa Radensky, Justin Jia, Kirielle Singarajah, Tom~M Mitchell, and Brad~A Myers.
\newblock Pumice: A multi-modal agent that learns concepts and conditionals from natural language and demonstrations.
\newblock In {\em Proceedings of the 32nd annual ACM symposium on user interface software and technology}, pages 577--589, 2019.

\bibitem[\protect\citeauthoryear{Li \bgroup \em et al.\egroup }{2020}]{li2020mapping}
Yang Li, Jiacong He, Xin Zhou, Yuan Zhang, and Jason Baldridge.
\newblock Mapping natural language instructions to mobile ui action sequences.
\newblock {\em arXiv preprint arXiv:2005.03776}, 2020.

\bibitem[\protect\citeauthoryear{Little \bgroup \em et al.\egroup }{2007}]{little2007koala}
Greg Little, Tessa~A Lau, Allen Cypher, James Lin, Eben~M Haber, and Eser Kandogan.
\newblock Koala: capture, share, automate, personalize business processes on the web.
\newblock In {\em Proceedings of the SIGCHI conference on Human factors in computing systems}, pages 943--946, 2007.

\bibitem[\protect\citeauthoryear{Luo \bgroup \em et al.\egroup }{2021}]{luo2021characteristics}
Yajing Luo, Peng Liang, Chong Wang, Mojtaba Shahin, and Jing Zhan.
\newblock Characteristics and challenges of low-code development: the practitioners' perspective.
\newblock In {\em Proceedings of the 15th ACM/IEEE international symposium on empirical software engineering and measurement (ESEM)}, pages 1--11, 2021.

\bibitem[\protect\citeauthoryear{OpenAI}{2024}]{chatgpt}
OpenAI. (2024). ChatGPT~(3.5) OpenAI.
\newblock \url{https://chat.openai.com}, 2024.

\bibitem[\protect\citeauthoryear{Pu \bgroup \em et al.\egroup }{2022}]{pu2022semanticon}
Kevin Pu, Rainey Fu, Rui Dong, Xinyu Wang, Yan Chen, and Tovi Grossman.
\newblock Semanticon: Specifying content-based semantic conditions for web automation programs.
\newblock In {\em Proceedings of the 35th Annual ACM Symposium on User Interface Software and Technology}, pages 1--16, 2022.

\bibitem[\protect\citeauthoryear{Pu \bgroup \em et al.\egroup }{2023}]{pu2023dilogic}
Kevin Pu, Jim Yang, Angel Yuan, Minyi Ma, Rui Dong, Xinyu Wang, Yan Chen, and Tovi Grossman.
\newblock Dilogics: Creating web automation programs with diverse logics.
\newblock In {\em Proceedings of the 36th Annual ACM Symposium on User Interface Software and Technology, arXiv:2308.05828}, 2023.

\bibitem[\protect\citeauthoryear{Rokis and Kirikova}{2022}]{Rokis2022challenges}
Karlis Rokis and Marite Kirikova.
\newblock Challenges of low-code/no-code software development: A literature review.
\newblock In {\={E}}rika Nazaruka, Kurt Sandkuhl, and Ulf Seigerroth, editors, {\em Perspectives in Business Informatics Research}, pages 3--17, Cham, 2022. Springer International Publishing.

\bibitem[\protect\citeauthoryear{Sahay \bgroup \em et al.\egroup }{2020}]{sahay2020supporting}
Apurvanand Sahay, Arsene Indamutsa, Davide Di~Ruscio, and Alfonso Pierantonio.
\newblock Supporting the understanding and comparison of low-code development platforms.
\newblock In {\em 2020 46th Euromicro Conference on Software Engineering and Advanced Applications (SEAA)}, pages 171--178. IEEE, 2020.

\bibitem[\protect\citeauthoryear{Schwartz \bgroup \em et al.\egroup }{2023}]{schwartz2023enhancing}
Sivan Schwartz, Avi Yaeli, and Segev Shlomov.
\newblock Enhancing trust in llm-based ai automation agents: New considerations and future challenges, 2023.

\bibitem[\protect\citeauthoryear{Sereshkeh \bgroup \em et al.\egroup }{2020}]{sereshkeh2020vasta}
Alborz~Rezazadeh Sereshkeh, Gary Leung, Krish Perumal, Caleb Phillips, Minfan Zhang, Afsaneh Fazly, and Iqbal Mohomed.
\newblock Vasta: a vision and language-assisted smartphone task automation system.
\newblock In {\em Proceedings of the 25th international conference on intelligent user interfaces}, pages 22--32, 2020.

\bibitem[\protect\citeauthoryear{Sugiura and Koseki}{1998}]{sugiura1998internet}
Atsushi Sugiura and Yoshiyuki Koseki.
\newblock Internet scrapbook: automating web browsing tasks by demonstration.
\newblock In {\em Proceedings of the 11th annual ACM symposium on User interface software and technology}, pages 9--18, 1998.

\bibitem[\protect\citeauthoryear{Touvron \bgroup \em et al.\egroup }{2023}]{touvron2023llama}
Hugo Touvron, Louis Martin, Kevin Stone, Peter Albert, Amjad Almahairi, Yasmine Babaei, Nikolay Bashlykov, Soumya Batra, Prajjwal Bhargava, Shruti Bhosale, et~al.
\newblock Llama 2: Open foundation and fine-tuned chat models.
\newblock {\em arXiv preprint arXiv:2307.09288}, 2023.

\bibitem[\protect\citeauthoryear{UiPath}{2023}]{uipath}
UiPath.
\newblock \url{https://www.uipath.com/}, 2023.

\bibitem[\protect\citeauthoryear{Wang \bgroup \em et al.\egroup }{2023}]{wang2023enabling}
Bryan Wang, Gang Li, and Yang Li.
\newblock Enabling conversational interaction with mobile ui using large language models.
\newblock In {\em Proceedings of the 2023 CHI Conference on Human Factors in Computing Systems}, pages 1--17, 2023.

\bibitem[\protect\citeauthoryear{Wang \bgroup \em et al.\egroup }{2024}]{wang2024mobile}
Junyang Wang, Haiyang Xu, Jiabo Ye, Ming Yan, Weizhou Shen, Ji~Zhang, Fei Huang, and Jitao Sang.
\newblock Mobile-agent: Autonomous multi-modal mobile device agent with visual perception.
\newblock {\em arXiv preprint arXiv:2401.16158}, 2024.

\bibitem[\protect\citeauthoryear{Zapier}{Accessed October 2023}]{zapier}
Zapier.
\newblock \url{https://zapier.com/}, Accessed: October, 2023.

\bibitem[\protect\citeauthoryear{Zheng \bgroup \em et al.\egroup }{2024}]{zheng2024gpt}
Boyuan Zheng, Boyu Gou, Jihyung Kil, Huan Sun, and Yu~Su.
\newblock Gpt-4v (ision) is a generalist web agent, if grounded.
\newblock {\em arXiv preprint arXiv:2401.01614}, 2024.

\end{thebibliography}


\onecolumn

\appendix
\section{Supplementary Material}

This supplementary material includes the following additional materials on IDA:
\begin{itemize}
    \item A demonstration video of IDA is available for further insight at \textbf{\href{http://tinyurl.com/ida-demo}{tinyurl.com/ida-demo}}.
    \item Overall description of IDA workflow in Section \ref{sec:IDA_workflow}.
    \item Section \ref{sec:IDA_screenshots} contains several screenshot examples for different workflow stages.
    \item Examples of a prompt, prompt input and prompt output for LLM approach are provided in Section .\ref{sec:prompt}.
    \item Section \ref{sec:surveys} describes pre-session and post-session user surveys.
\end{itemize}
 
\subsection{IDA Workflow Pipeline}

\label{sec:IDA_workflow}

\begin{figure*}[htbp]
\centering
\includegraphics[width=\textwidth]{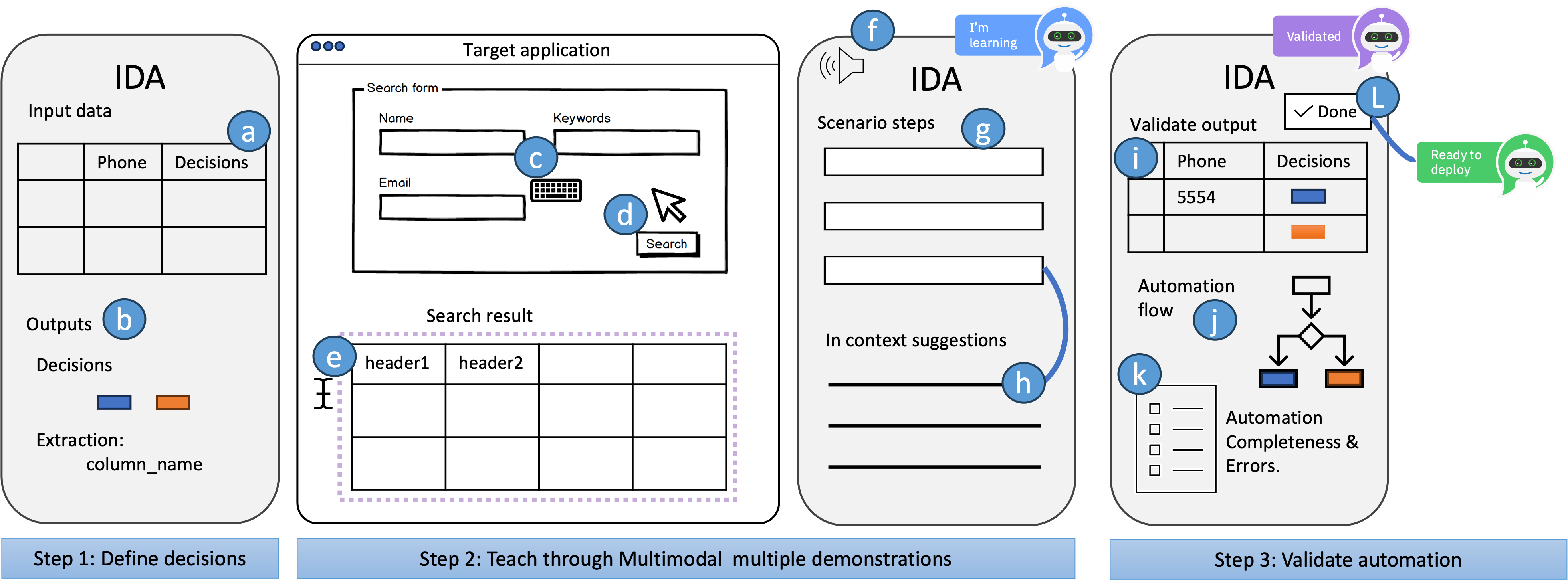}
\caption{\textbf{The workflow of IDA}. There are three steps to creating a web automation with IDA. (Step 1: Define) the user defines the sample input data (a) and outputs (b) for the automation.  (Step 2: Teach) to teach IDA how to perform scenarios, the user can demonstrate via keyboard (c) and mouse (d) on the application and IDA UI presents the captured scenario steps (g) with auditory feedback (f) which the user can also edit.  To support conditional execution, the user can select a semantic UI element (e) and configure a conditional by either specifying the condition in voice, or by selecting a recommended action in the IDA UI (h).(Step 3: Validate) IDA UI presents the full automation map from multiple demonstrations (j) so the user can confirm the complete specification. It also recommends to the user on additional scenarios that should be demonstrated on uncovered decisions or condition space.  It further presents issues discovered during the synthesis of the demonstrations into a single automation program (k) The user requests IDA to show how it is performing the automation on additional examples of the data (i).  The user can see a near real-time view of how the program is progressing along the program map as well as within the application (j).  When the program is validated, the user can mark the automation as ready to deploy (L) .}
\label{fig:teaserfigure}
\end{figure*}

\newpage
\subsection{Examples of IDA screenshots}

\label{sec:IDA_screenshots}

\subsubsection{Define stage}
\begin{figure}[ht!]
\centering
    \includegraphics[width=0.4\textwidth]{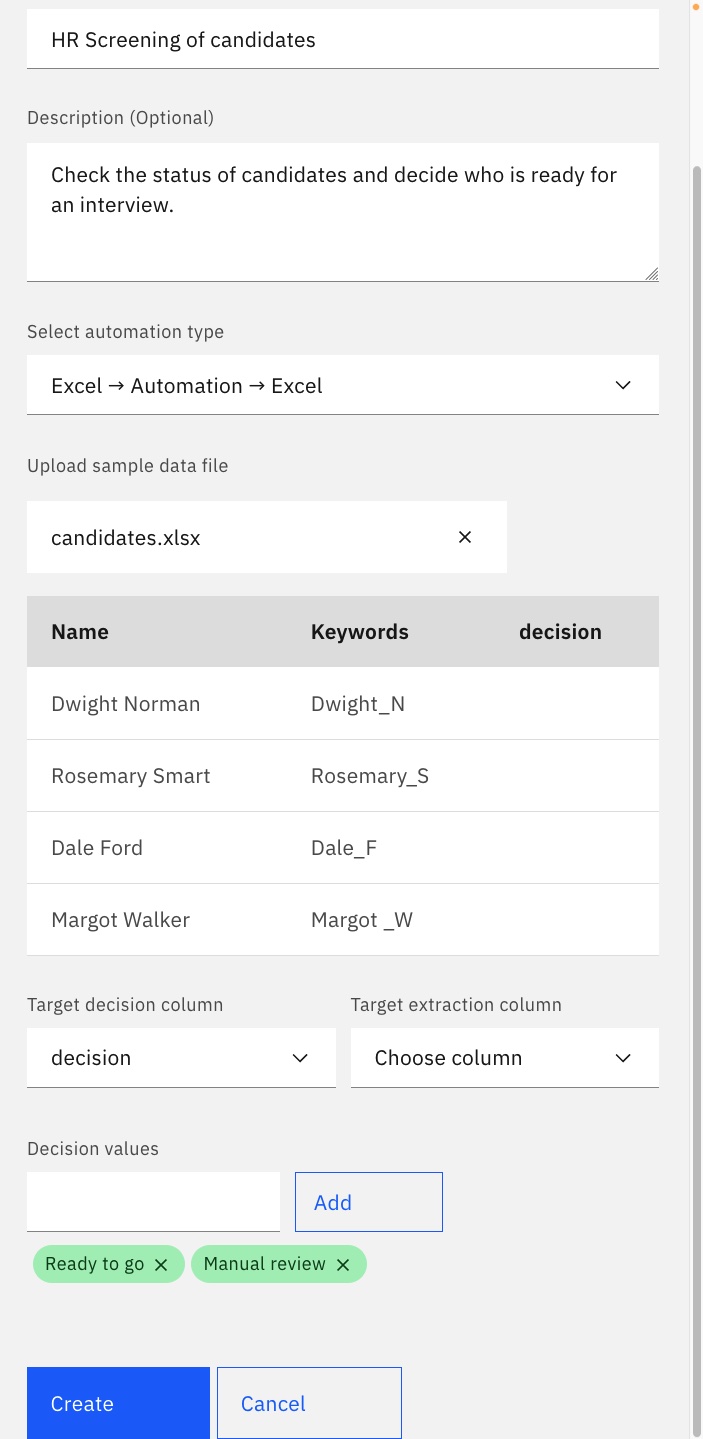}
    \caption{Creating a new automation for the HR Candidate Screening process as part of the  \textit{define} stage. 
    Cassie specifies the external inputs and outputs for a new automation (\textit{Automation Name} and \textit{Description}).  She selects the automation type in order to configure the inputs and outputs to the automation \textit{Excel $\rightarrow$ Automation $\rightarrow$ Excel}.  This maps to an automation template that iterates over an input Excel file and invokes the automation for each row, as well as write back to this Excel file. Cassie uploads sample data from a representative Excel file which she wants to use for teaching IDA. The \textit{Target extraction column} is used to define the output location of the scraped data in the Excel file.  The \textit{Target decision column} is used to define the column in Excel for writing decision values.  The decisions \textit{Ready to go} and \textit{Manual review} are defined as possible decision values.   
    }
    \label{fig:define_with_values}
\end{figure}


\newpage

\subsubsection{Teaching stage: Guiding Cassie to demonstrate multiple scenarios}
\begin{figure}[h!]
    \centering
    \includegraphics[width=0.5\textwidth]{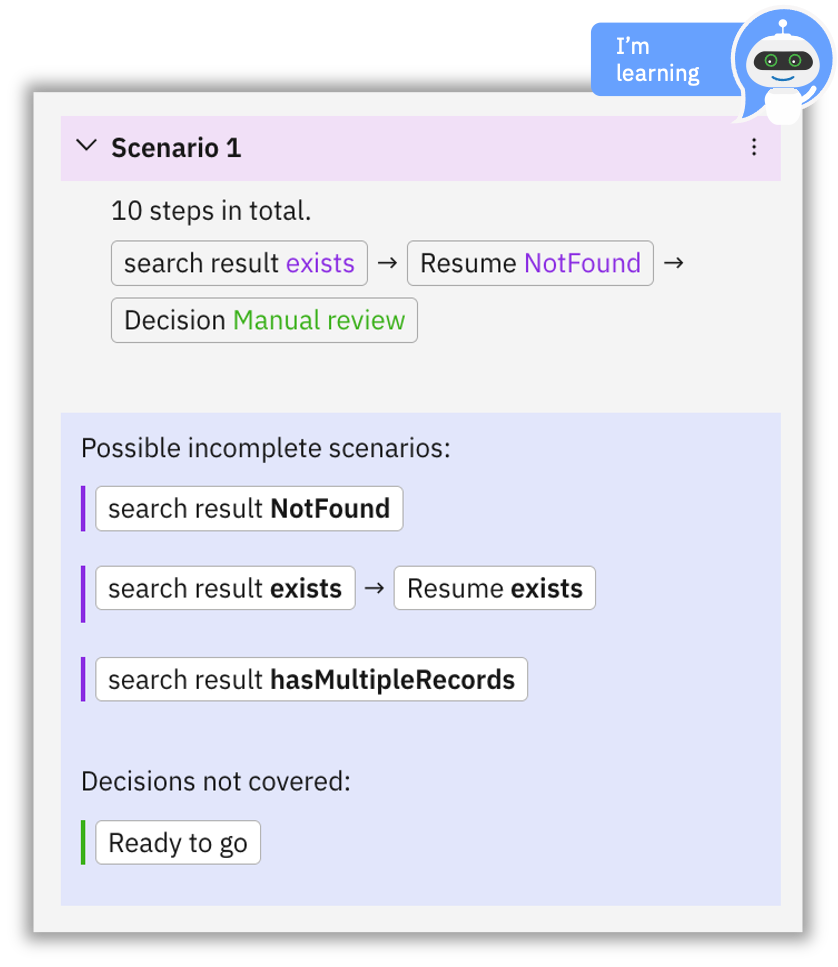}
    \caption{Recommending missing scenarios and conflicts. After Cassie completes demonstrating a scenario, IDA presents a summary of the captured scenario and provides recommendations for additional scenarios that Cassie should consider to demonstrate.  These scenarios correspond to uncovered values in the decision space and uncovered conditional states in the semantic objects [DG1, DG3]. IDA suggests demonstrating the ready-to-go which was specified during the Define stage. IDA also suggests Cassie to demonstrate the \textit{search result exists $\rightarrow$ Resume NotFound} scenario based on the coverage analysis of the semantic elements and the conditions that were demonstrated so far.  This serves as a guide and reminder for Cassie to start recording the second scenario. When Cassie starts recording the second scenario, she must be consistent with the previous scenario and record the same actions leading to a shared branching point from the first demonstrated scenario. 
    For example, for the manual-review2 Cassie should demonstrate the same initial steps up to the condition on the search results, where she can continue to record the search results NotFound steps.  Similarly for the ready-to-go scenario, Cassie should demonstrate the same steps up to the final decision on the resume element.  If a step recorded by IDA is inconsistent with the program so far, the scenario will be marked with an error.}
    \label{fig:teach_reminder}
\end{figure}
%

\newpage
\subsubsection{Teaching stage: Visualizing the full automation program} 
\begin{figure}[h!]
    \centering
    \includegraphics[width=0.5\textwidth]{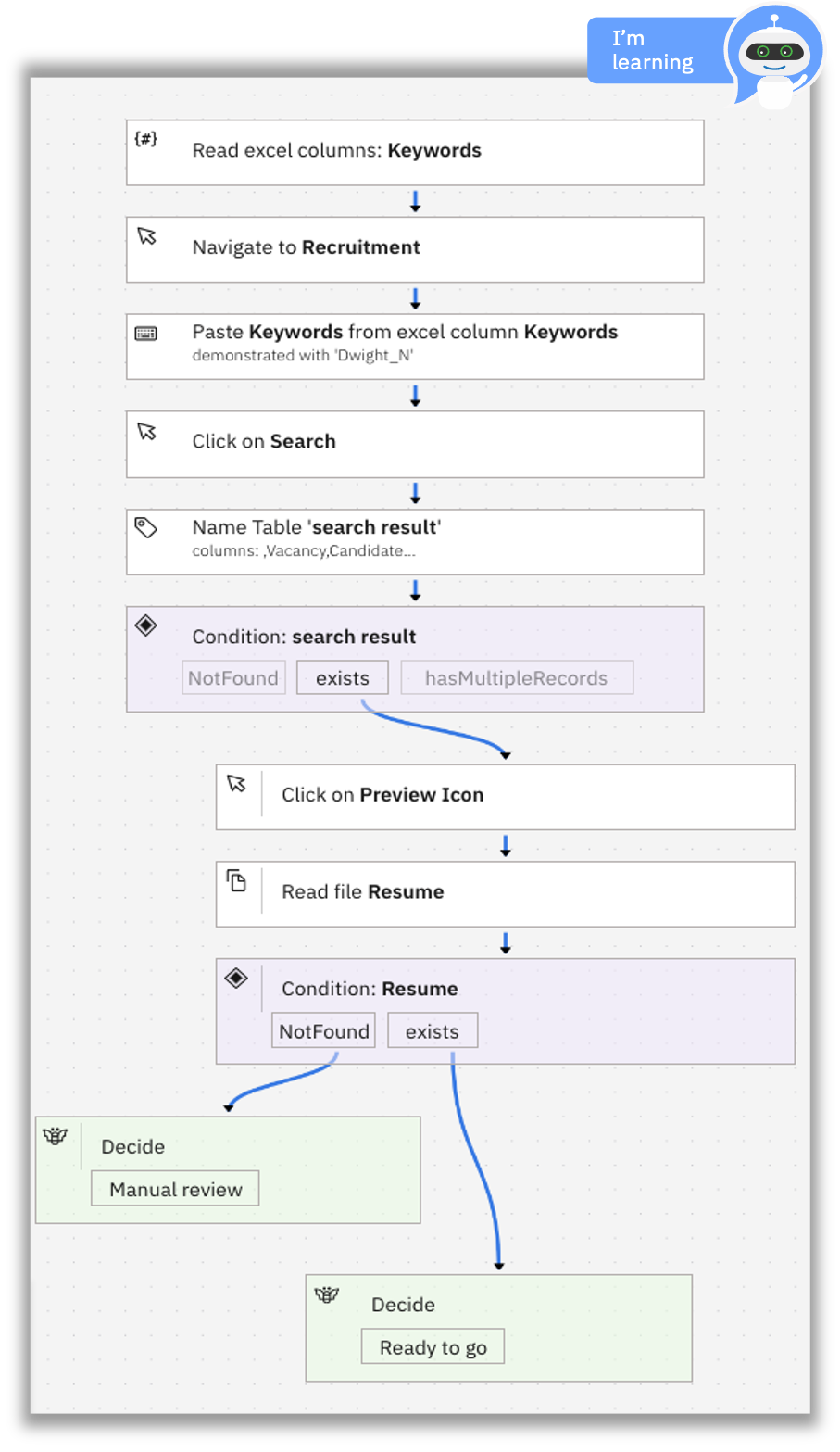}
    \caption{\textbf{Automation map view}: after demonstrating the manual-review1 and ready-to-go scenarios, IDA synthesizes the program and presents a visual representation of the program as a whole. The automation map allows Cassie to review the entire automation, and Cassie can observe that when the "resume" exists, the decision is "Ready to go" whereas if it does not exist, the decision is "manual review."}
    \label{fig:teach_automation_map}\end{figure}
%


\newpage

\subsection{LLM Tasks: Prompt example}

\label{sec:prompt}

\begin{figure}[h!]
    \centering
    \includegraphics[width=0.8\textwidth]{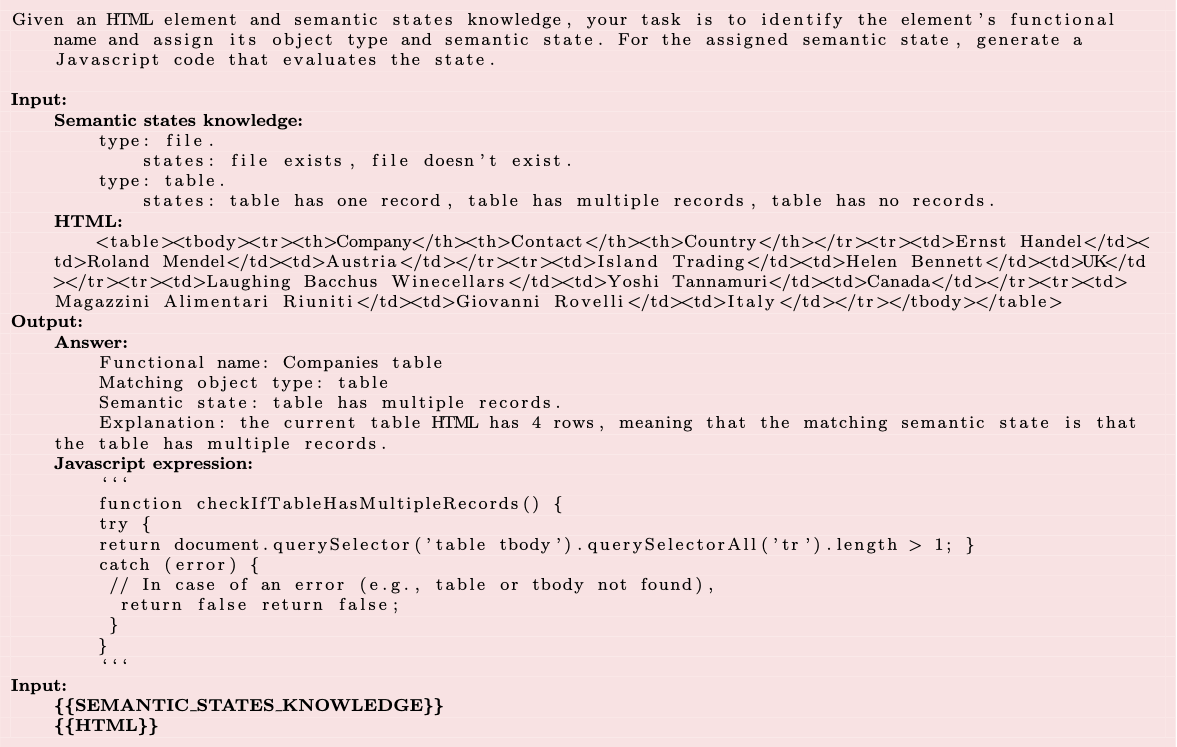}
    \caption{A one-shot LLM prompt for semantic object detection. An HTML element and possible states are given, where the task is to identify the functionality of the element and to generate a Javascript code that evaluates the state of the element.}
    \label{fig:prompt_example}
\end{figure}


\begin{figure}[h!]
    \centering
    \includegraphics[width=0.8\textwidth]{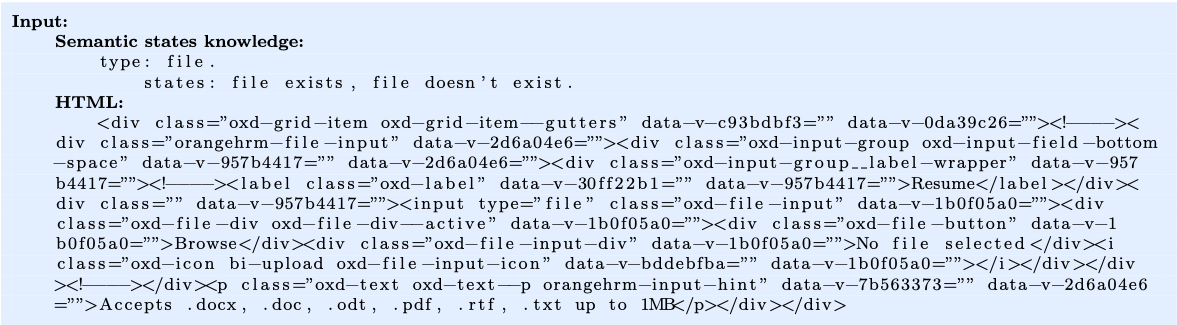}
    \caption{An example of the LLM prompt input, semantic states knowledge and HTML.}
    \label{fig:input_example}
\end{figure}    

\begin{figure}[h!]
    \centering
    \includegraphics[width=0.8\textwidth]{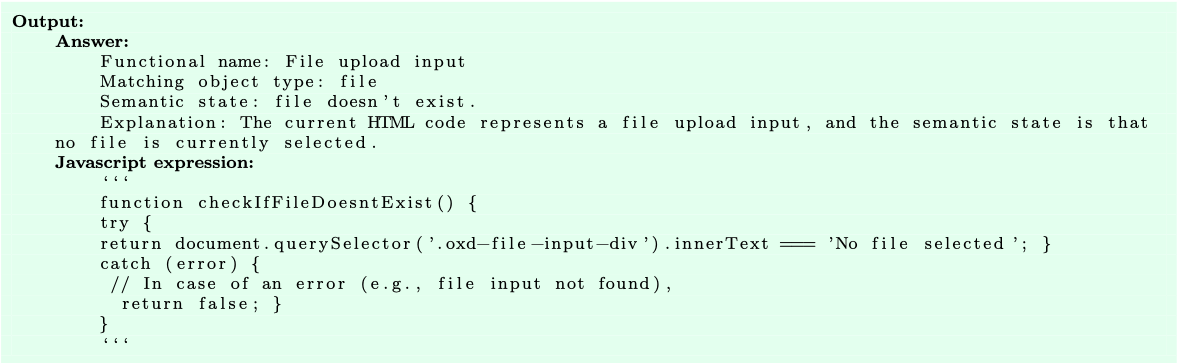}
    \caption{An example of the LLM prompt output generated by Llama2 70B.}
    \label{fig:output_example}
\end{figure}

\newpage

\subsection{User surveys}

\label{sec:surveys}
Our study is approved by the ethics review board at [anonymous].
Each participant in the user study was requested to complete two surveys. The first survey, conducted prior to the session, aimed to obtain informed consent to participation and gather background information, including demographics, experience, and skills. The second survey, administered after the session, aimed to capture users' feedback regarding the experience they had just undergone.

\subsubsection{Pre-session survey}
The pre-session survey included the following questions.
\begin{itemize}
    \item Informed Consent;
    \item Name;
    \item Education;
    \item What is the academic discipline of your Bachelor’s degree?
    \item Which domain/department do you work at (HR, Procurement, Admin, Finance, \ldots)?
    \item What is your role at the organization?
    \item Do you have an experience in programming?
    \item If you answered "yes", number of years? Have you created in the past a bot for automation purposes?
    \item Have you ever used automation tools?
    \item If you answered "Yes" - which tools?
\end{itemize}

\subsubsection{Post-session survey} 
\mbox{}
\label{ques2}

Eleven questions of post-session survey used Likert scale rating, where subjects were asked to indicate their level of agreement or disagreement with the following statements using a scale ranging from "strongly agree" to "strongly disagree."

\begin{itemize}
    \item I found the system easy and convenient to use.
    \item It was simple for me to understand where everything I needed was on the screen.
    \item I think I would need help from a technical person to use a system like IDA.
    \item If I made a mistake, it was easy for me to fix it and continue using the feature.
    \item I felt that performing the tasks required a lot of effort from me.
    \item The use of audio helped me complete the task.
    \item I think the system was too complicated to use.
    \item I think I could manage on my own with a system like that.
    \item If I had access, I would use IDA to automate my everyday work.
    \item The use of audio didn't make a difference in how well I completed the task.
    \item I felt that the tasks didn't require a lot of attention from me.
\end{itemize}

In addition, the users answered four open-ended questions:
\begin{itemize}
    \item Which elements of the system were the least convenient for you? If you have an idea how to change them so that they are more convenient, please specify.
    \item What were the most convenient elements in the system for you? Do you have an idea how to improve these points even more?
    \item Do you feel that you would trust this system to perform actions for you on a daily basis?
    \item Can you think of cases where you could use a system like this in your professional world? Give examples.
\end{itemize}

\end{document}